\documentclass[aps,prl,longbibliography,floatfix,reprint]{revtex4-1}
\usepackage{amsmath}
\usepackage{graphicx}
\usepackage{amsfonts}
\usepackage{color}
\usepackage{placeins}

\begin{document}

\title{Resistivity of high pressure phosphorus phases}

\author {Xinyu Li}
\affiliation{Department of Mechanical Engineering, University of Texas, Austin, Texas 78712-1591, USA}
\author{ Philip B. Allen }
\email{philip.allen@stonybrook.edu}
\affiliation{ Department of Physics and Astronomy,
              Stony Brook University, 
              Stony Brook, New York 11794-3800, USA }

\date{\today}

\begin{abstract}

Simple cubic (sc) ``black phosphorus'' (denoted ``BP''), stable at $P>$10GPa, seems an ordinary 
metal.  It has electron-phonon-driven superconductivity at $T_c \approx$ 5-10 K.
The  A17 phase, stable at atmospheric pressure, has a narrow gap, becomes semimetallic at $P$=1 GPa,
and has a smooth transition to topological metal behavior at $P \approx$5 GPa.  
The A7 phase, stable for $5<P<10$ GPa, is metallic, superconducting, and less conventional than the sc phase.
Some insights are extracted from analysis of resistivity $\rho(T)$ at various pressures.  A surprising order-of-magnitude
disagreement between theory and experiment is discussed.

\end{abstract}

\maketitle

\section{Introduction}

Black phosphorus (BP) becomes metallic under pressure ($P$) \cite{Bridgman1914,Jamieson1963}.  The absolute resistivity
was recently measured by Li {\it et al.} \cite{Li2018} at pressures up to 15 GPa, from $T=$1.5K to 300K. 
The simple cubic phase (stable from $P$=10 to 137 GPa \cite{Akahama1999}) seems a normal metal, 
and superconducts, at least in the lower $P$ range, at $T\le10$K.
Analysis of resistivity, with input from density-functional (DFT) band theory, provides a way to 
extract the electron-phonon coupling constant $\lambda$ \cite{Poole2000,Allen1987}.

\begin{figure}
\includegraphics[angle=0,width=0.42\textwidth]{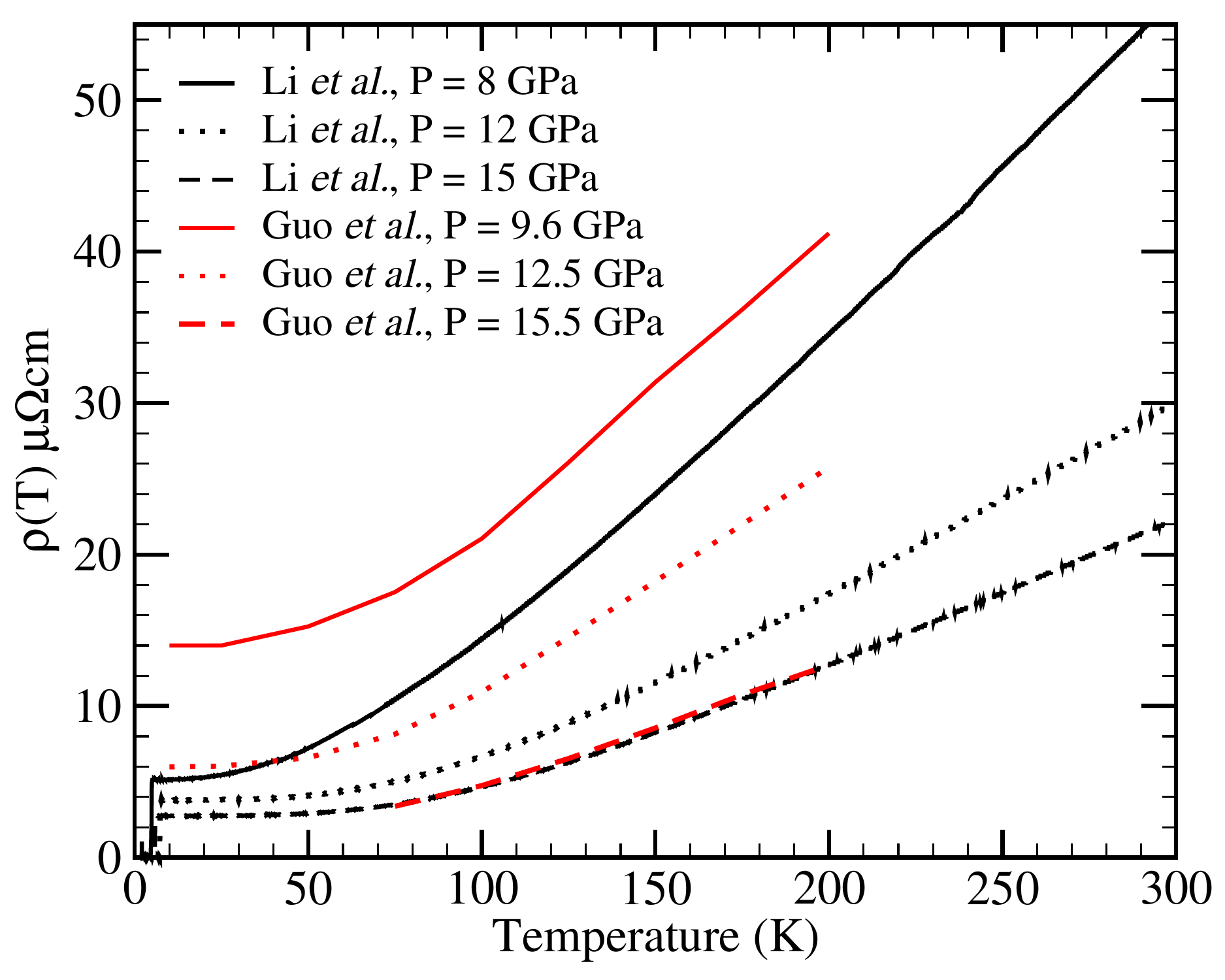}
\caption{\label{fig:rho} Absolute resistivities measured at $0<T<300$K by Li {\it et al.} \cite{Li2018} for sc structure BP 
are compared with relative resistivities measured by Guo {\it et al.} \cite{Guo2017} for $0<T<200$K.  The relative resisitivity of the
15.5 GPa case was scaled up by a factor 25.2 to give agreement with the absolute resistivity at 15 GPa.
The data then suggest that Guo's sample had a higher residual resistivity $\rho_0$ than Li's, and a subtraction 
$\Delta\rho_0=-2.74 \ \mu\Omega$cm was made to line it up with Li's.  The same scaling factor 25.2 
(with a scaled version of $\Delta\rho_0$) was done for Guo's data at two other pressures.}
\end{figure}

\begin{figure}
\includegraphics[angle=0,width=0.42\textwidth]{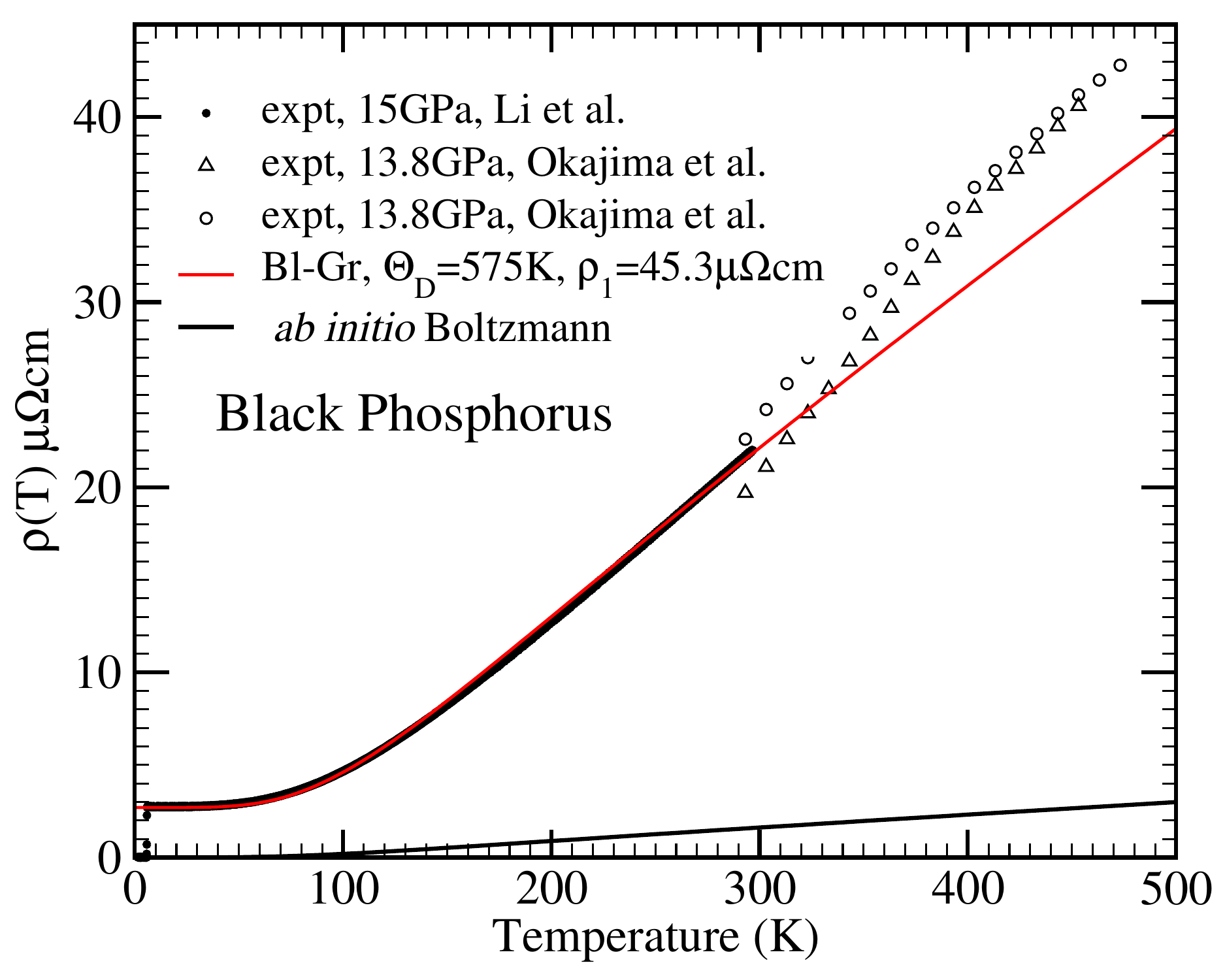}
\caption{\label{fig:14} The 15 GPa data of Li {\it et al.} \cite{Li2018} for $T<300$K, and a Bloch-Gr\"uneisen fit (up to 500K).  
Also shown are data from Okajima {\it et al.} \cite{Okajima1984} taken
at $P=$13.8 GPa for $T$ going from room temperature up to 473K.  The circles are measurements taken as $T$ increases,
and triangles are data as $T$ decreases.  The difference between these data presumably arises from difficulties
such as maintaining a constant pressure while relieving anisotropic strain.
The curve labeled ``{\it ab initio} Boltzmann''  is from a different variational solution,
Eq. \ref{eq:var}, of the Bloch-Boltzmann equation,
using the {\it ab initio} $\alpha^2F$ from Ref. \onlinecite{Wu2018} at $P$=17.5 GPa, and the free electron
$(n/m)_{\rm eff}$ value for 15 GPa.      
The discrepancy of a factor 10 is not understood.}
\end{figure}

Our analysis for black phosphorus, reported here, indicates that this goes seriously wrong.  Therefore it
is important to ask how reliable are resistivity measurements at such pressures.  The paper by Guo {\it et al.}
\cite{Guo2017} gives relative resistivities at various pressures.  These are compared with the absolute
resistivites of Li {\it et al.} in Fig. \ref{fig:rho}.  There is general agreement about the shape of the $T$ dependence.
Higher $T$ measurements at 13.8 GPa were published by Okajima {\it et al.} \cite{Okajima1984}.  These
are shown in Fig. \ref{fig:14}.  The data are encouragingly similar
to the recent results of Li {\it et al.} \cite{Li2018}.
One can tentatively accept Li's data as a realistic standard.

\par
\par
\par
\par

The Bloch-Gr\"uneisen formula \cite{Bloch1929,Ziman1960} for electrical resistivity of a metal is
a variational approximation for the solution of the Bloch Boltzmann equation \cite{Bloch1929} for electrons 
scattering from phonons in metals.  It uses a Debye model.  The formula is
\begin{equation}
\rho_{\rm BG}(T)=\rho_0 + \rho_1 f_{\rm BG}(T/\Theta_D)
\label{eq:BG}
\end{equation}
where $\rho_0$ is the residual resistivity from impurity scattering, $\Theta_D$ is the Debye 
temperature, and $\rho_1$ is given by
\begin{equation}
\frac{1}{\rho_1}=\left(\frac{n}{m}\right)_{\rm eff} e^2 \tau_D.
\label{eq:rho1}
\end{equation}
The factor $(n/m)_{\rm eff}$ from DFT computations will be explained shortly.
The term $\tau_D$ is a normalization factor for the scattering lifetime $\tau_{\rm tr}(T)$;
$1/\tau_{\rm tr}$ is $1/\tau_D$ times a dimensionless factor that,
in Bloch-Gr\"uneisen theory is $f_{\rm BG}(T/\Theta_D)$.  The equations are
\begin{equation}
\frac{1}{\tau_D}=\frac{2\pi\lambda_{\rm tr}k_B \Theta_D}{\hbar},
\label{eq:tau}
\end{equation}
where $\lambda_{\rm tr}$ is the transport version of the dimensionless electron-phonon coupling constant
$\lambda$.  The dimensionless function $f_{\rm BG}(T/\Theta_D)=f_{\rm BG}(y)$ is
\begin{equation}
f_{\rm BG}(y)=4y\int_0^1 dx x^3 \left(\frac{x/2y}{\sinh(x/2y)}\right)^2.
\label{eq:BG}
\end{equation}
This assumes a three-dimensional Debye spectrum for the phonons.  At large $y=T/\Theta_D$, the factor
in parentheses in Eq. \ref{eq:BG} goes to 1 and $f_{\rm BG}\rightarrow T/\Theta_D$, giving
the familiar high $T$ linear resistivity.  At low $T$, the function $f_{\rm BG}$ becomes $64(T/\Theta_D)^5$ times
the integral $\int_0^\infty dz z^5 /\sinh^2 z = 15\zeta(5)/2$.  This gives the familiar $T^5$ temperature dependence.
Finally, the factor $(n/m)_{\rm eff}$ is 
\begin{equation}
\left(\frac{n}{m}\right)_{\rm eff}=\frac{1}{V}\sum_k \frac{1}{\hbar^2}\frac{\partial^2 \epsilon_k}{\partial k_x^2}f_k
=\frac{1}{V}\sum_k v_{kx}^2\left(-\frac{\partial f_k}{\partial\epsilon_k}\right).
\label{eq:}
\end{equation}
The index $k$ is short for $(\vec{k}n\sigma)$, the wavevector, band index, and spin needed to label a state.
The derivative $-\partial f/\partial \epsilon$ of the Fermi function is accurately replaced by a delta function,
$\delta(\epsilon-\mu)$, so $(n/m)_{\rm eff}=N(\epsilon_F)\langle v_x^2 \rangle$, where $N(\epsilon_F)$ is the density of states
(per unit volume) at the Fermi level $\epsilon_F=\mu$, and 
$\langle v_x^2 \rangle=\sum_k v_k^2 \delta(\epsilon_k-\mu)/\sum_k \delta(\epsilon_k-\mu)$
is the Fermi surface average of the squared $x$ component
of the electron's group velocity.  In an anisotropic material, a first guess would be that the conductivity tensor
$\sigma_{\alpha\beta}$ is given by the same formula, except $\langle v_x^2 \rangle$ is replaced by 
$\langle v_\alpha v_\beta \rangle$.  The older terminology ``optical mass'' is still sometimes used for the
mass in the denominator of $(n/m)_{\rm eff}$.  The reason for abandoning this terminology is that $n$ and $m$
are not separately definable except in semiconductors with small carrier densities in a parabolic band.  For phosphorus,
the choice $n=5$ electrons per atom works fairly well for the sc phase, but not so well for lower $P$ phases, or for
high $P$ phases where $3d$ electron states start to be occupied.


A more complete theory is also available, using a variational solution of the Bloch-Boltzmann
equation \cite{Allen1971} which avoids the Deybe approximation of Bloch-Gr\"uneisen theory:
\begin{eqnarray}
\rho(T)&\approx& \rho_0+ \frac{1}{(n/m)_{\rm eff}e^2\tau(T)}; \\ \nonumber
\frac{\hbar}{\tau(T)}&=& 4\pi k_B T \int_0^\infty \frac{d\omega}{\omega}\alpha_{\rm tr}^2 F(\omega) 
\left[ \frac{\hbar\omega/2k_B T}{\sinh(\hbar\omega/2k_B T)} \right]^2.
\label{eq:var}
\end{eqnarray}
Here $\alpha_{\rm tr}^2 F$ is a modified version \cite{Allen1971} of the function $\alpha^2 F$ used in Eliashberg theory
of electron-phonon superconductors \cite{Eliashberg1960,Parks1969}.  The additional information in
$\alpha_{\rm tr}F$ will not overcome the discrepancy between
theory and experiment.

Mass renormalization by interactions is another worry.  Coulomb renormalization is usually well
incorporated in the DFT band masses.  Electron-phonon renormalization is seen in ac conductivity \cite{Allen2015},
$\sigma(\omega)\approx(n/m)_{\rm eff}e^2/(1/\tau+i\omega\Lambda(\omega))$, but drops out in the dc limit.

\section{fitting experimental resistivity}

\par
\begin{figure}
\includegraphics[angle=0,width=0.42\textwidth]{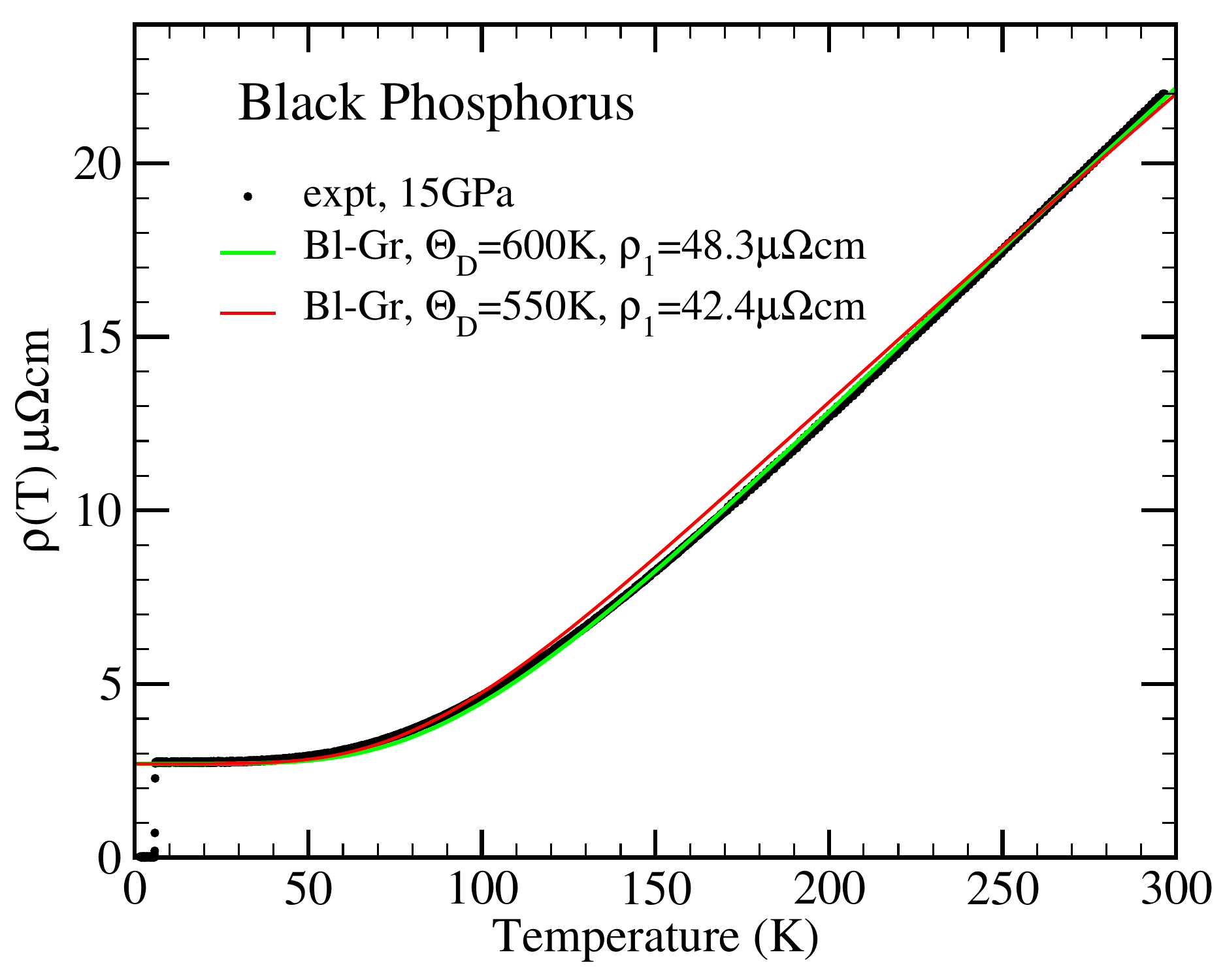}
\caption{\label{fig:15} At 15GPa, the resistivity \cite{Li2018} of sc structure BP is fitted well by a Bloch-Gr\"uneisen formula,
with $\Theta_D\approx$575 K.  If the electron-phonon coupling strength is taken as $\lambda_{\rm tr}=0.8$,
then the Drude plasma frequency $\omega_p$ needs to be 6.4 eV.}
\end{figure}
\par

Resistivity measurements by Li {\it et al.} \cite{Li2018} cover $0<T<300$K, and $P$ up to
15 GPa.  The sc phase at $P>10$ GPa is the most conventional. Resistivity
$\rho(T)$, and Bloch-Gr\"uneisen fits, are shown in Fig. \ref{fig:15}.  The fits
is as good as normally expected.  An even better fit is shown in Fig. \ref{fig:14}.  In principle there should be
deviations from Bloch-Gr\"uneisen because of deviations of the phonon spectrum from Debye.  The
deviations are particularly small, probably because the sc crystal structure has simple and rather Debye-like phonons.

\begin{figure}
\includegraphics[angle=0,width=0.42\textwidth]{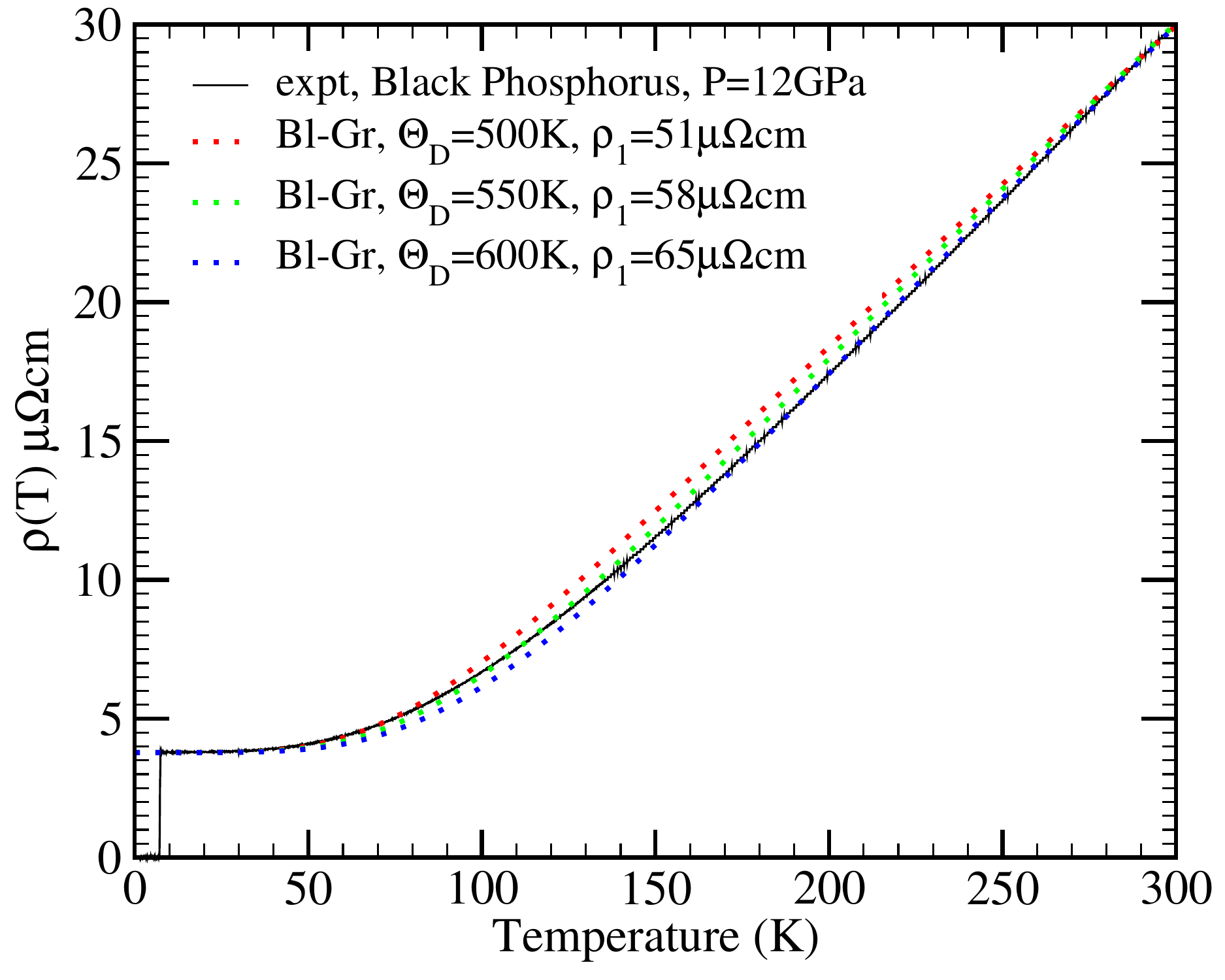}
\caption{\label{fig:12} At 12GPa, the structure of BP is still sc.  The resistivity \cite{Li2018}
is fitted less well by a Bloch-Gr\"uneisen formula than at 15GPa.}
\end{figure}

\begin{figure}
\includegraphics[angle=0,width=0.42\textwidth]{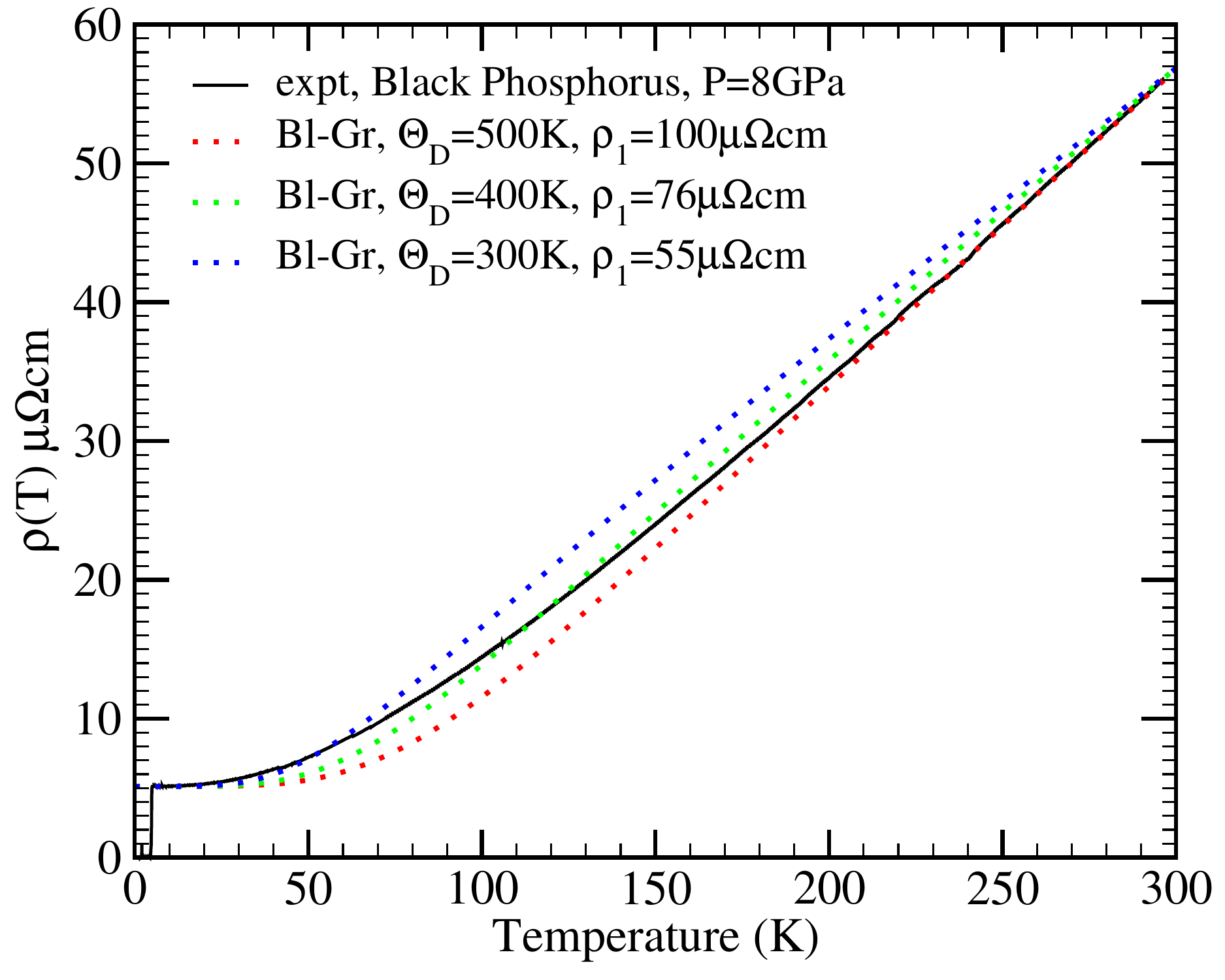}
\caption{\label{fig:8} At 8GPa, the structure of BP is A7.  The resistivity \cite{Li2018}
deviates significantly from a Bloch-Gr\"uneisen formula.}
\end{figure}

At 12 GPa, the Bloch-Gr\"uneisen fit, shown in Fig \ref{fig:12}, is not as good as normally expected, but not totally bad.
In the A7 phase at 8 GPa, shown in Fig. \ref{fig:8}, the fit doesn't work.  
The choice $\Theta_D$=300K fits at low $T$ but fails at higher $T$.
The choice $\Theta_D$=500K fits near room temperature but fails at low $T$.  The choice $\Theta_D$=400K does not work
well at either end.
What is the reason?  Coulomb or impurity scattering would not help.  Chan {\it et al.} \cite{Chan2013} find the sc structure unstable
in harmonic approximation for $P<$20GPa.  This suggests the importance of anharmonicity
to dynamically stabilize the sc phase at 10GPa$<P<$20GPa.  This anharmonicity probably persists in the 
lower $P$ A7 phase.  Perhaps anharmonic phonon-phonon
interactions change or even invalidate the Boltzmann quasiparticle theory.  

\par
\par
\par
\par

\section{testing the interpretation}

The fitting factor $\rho_1$ used in Eq. \ref{eq:BG} can be written, using Eqs. \ref{eq:rho1} and \ref{eq:tau} as
\begin{equation}
\rho_1=\frac{2\pi \lambda_{\rm tr}k_B \Theta_D/\hbar}{\epsilon_0 \omega_p^2},
\label{eq:rat}
\end{equation}
where $\omega_p^2=(n/m)_{\rm eff}e^2/\epsilon_0$ is the square of the Drude plasma frequency.
For the $P$=15 GPa data of Li {\it et al.} \cite{Li2018}, 
values of $\rho_1\sim45 \ \mu\Omega$cm and $\Theta_D\sim 575$K give good fits to $\rho(T)$ data. 
This requires $\omega_p^2/\lambda_{\rm tr}$=51.5 (eV)$^2$. 
The value of $\lambda_{\rm tr} \approx \lambda$ can be estimated from the superconducting $T_c$ to be $\sim0.5-0.8$.
Chan {\it et al.} \cite{Chan2013} compute $\lambda\sim0.7-0.8$ and diminishing as $P$ increases, for $P>$20GPa.
Flores-Livas {\it et al.} \cite{Flores-Livas2017} and
Wu {\it et al} \cite{Wu2018} compute $\lambda\sim0.5-0.65$.
These numbers are well in line with the measured superconducting $T_c$'s.
If such values are used to fit $\rho(T)$ data, they
require a Drude plasma frequency $\omega_p\sim 5-6$ eV.
The  next section tests this by band calculations.
The results for the 15 GPa sc phase are similar to free electron values, with $\omega_p > 20$ eV.
Then $\lambda_{\rm tr}$ should be higher by $\sim$16, {\it i.e.} $\lambda\sim$10, an unphysically large value.

\section{Electronic structure calculations}

\begin{figure}
\includegraphics[angle=0,width=0.42\textwidth]{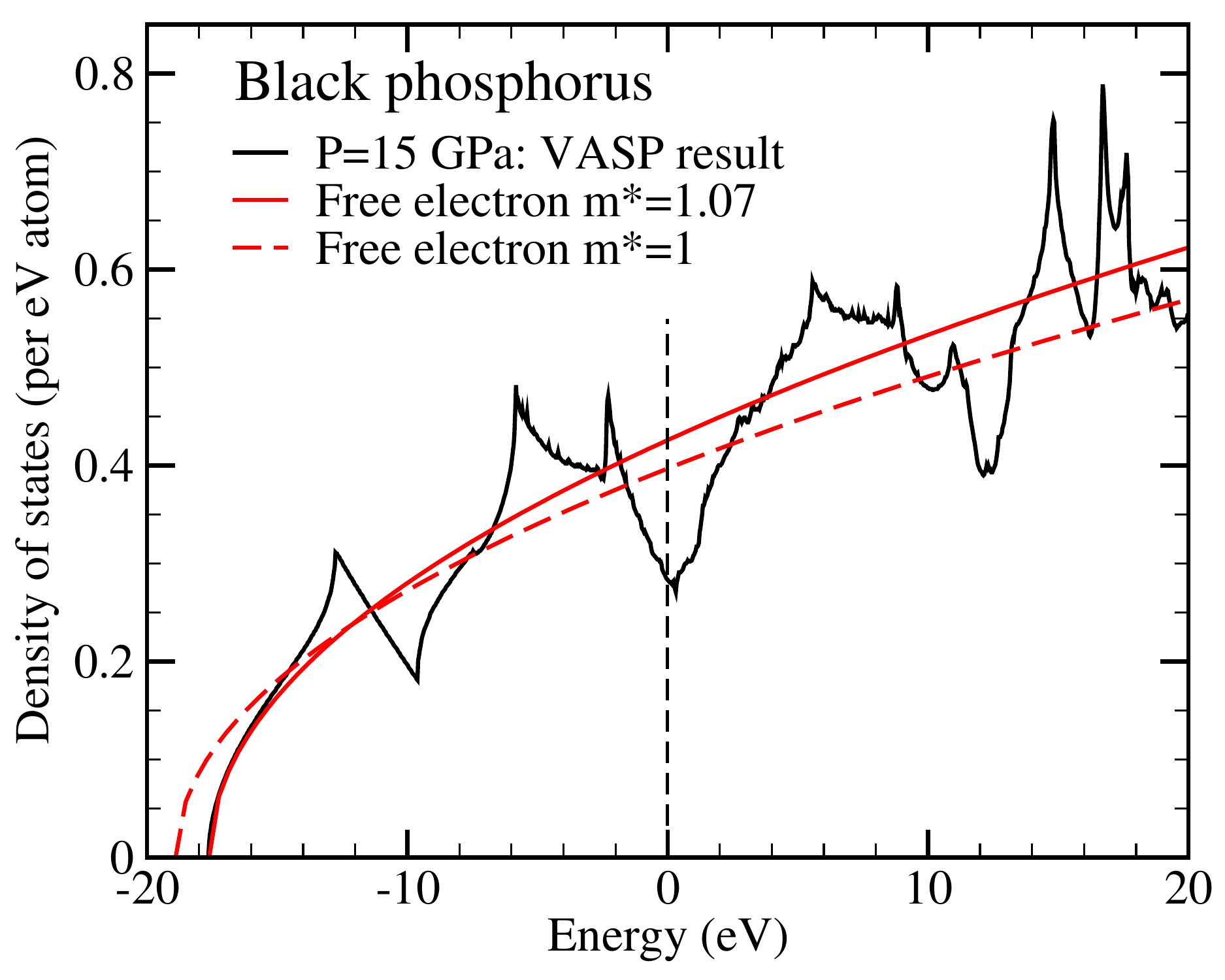}
\caption{\label{fig:dos} The DFT density of states from VASP agrees well on average with the free electron
form.  The solid red curve uses a free electron Fermi level with the electron mass increased by 1.07 in order
to coincide with the VASP value.  The dashed red curve uses  the free electron Fermi level and mass.
All curves then have the Fermi level set to zero.
The total number of states below the Fermi level is 5.}
\end{figure}

Electronic structure calculations have been done for the simple cubic phase by Aoki {\it et al.} \cite{Aoki1987},
Rajagopalan {\it et al.} \cite{Rajagopalan1989}, Chan {\it et al.} \cite{Chan2013}, 
Flores-Livas {\it et al.} \cite{Flores-Livas2017}, and Wu {\it et al.} \cite{Wu2018}.
Their results show that a free electron gas model describes the general features of the bands.
This is illustrated in Fig. \ref{fig:dos}.  Numerical results are in Table I.  These show
that sc black phosphorus is reasonably well modeled as a free electron gas.
The density of electrons is 5 per atom, with one atom per cell of lattice constant $a$.  Values of $a$ near  
$a=2.4\AA$ at $P\sim$15 GPa were measured by many authors \cite{Akahama1999,Kikegawa1983,Clark2010,Guo2017}.

Density functional theory (DFT) calculations were performed on the sc phase 
with the projector augmented wave method \cite{Blochl1994,Kresse1999}, 
as implemented in the Vienna Ab Initio simulation package VASP \cite{Kresse1994,Kresse1996}. 
A 47$\times$47$\times$47 $\vec{k}$-point mesh, a plane wave cutoff of 500 eV, and a force convergence tolerance 
of 2.5 meV/$\AA$ were employed in structural relaxation and density of states calculations. 
The sc phases were simulated under pressures of 12 and 15 GPa. The $(n/m)_{\rm eff}$ values 
were calculated using the code BoltzTrap \cite{Madsen2006}
based on band structure from VASP calculations.
\begin{table}
\begin{tabular}{| l | l | l |} \hline\hline
{\em parameter}&{\em free electrons}&{\em DFT} \\ \hline\hline
 $n$ &\multicolumn{2}{c  |}{5/$a^3=3.76\times10^{29}{\rm m}^{-3}$}  \\  \hline
$k_F$ & 2.23$\times 10^{10}{\rm m}^{-1}$ & \\ \hline
$r_s$ & 1.62 & \\ \hline
$<v_F^2>^{1/2}$ & 2.58$\times10^6$m/s&2.75$\times10^6$m/s \\ \hline
$\epsilon_F$ & 18.9 eV & 17.6 eV \\ \hline
$N(\epsilon_F)$ & 0.397/eV atom & 0.283/eV atom  \\ \hline
$(n/m)_{\rm eff}$ & 4.13$\times10^{59}$/kg m$^3$ & 3.35$\times10^{59}$/kg m$^3$ \\ \hline
$(n/m)_{\rm eff}e^2$ & 1.06$\times10^{22}/\Omega$ms & 0.86$\times10^{22}/\Omega$ms  \\ \hline
$\hbar\omega_p$ & 22.8 eV & 20.5 eV \\ \hline\hline
\end{tabular}
\caption{Theoretical parameters of sc phosphorus at 15GPa.  The lattice constant $a=2.369\AA$ is used.  
The dimensionless electron gas parameter $r_s$ is $ (9\pi/4)^{1/3}/a_B k_F$.
The Fermi energy $\epsilon_F$ is measured from the bottom of the $3s$ valence band.
The parameter $(n/m)_{\rm eff} $  is $N(\epsilon_F)<v_F^2>/3$. 
The Drude plasma frequency is $\omega_p =\sqrt((n/m)_{\rm eff}e^2/\epsilon_0)$.
The DFT parameters were computed using VASP (the first three) and BoltzTraP \cite{Madsen2006} (the last three).}
\end{table}

%
%

\par
\par

If we choose a reasonable value $\lambda_{\rm tr}\sim 0.8$, then the BoltzTraP result 
$\omega_p=$20.5 eV predicts at $T$=300K, using the high $T$ limit, 
resistivity $\rho(300{\rm K})-\rho(0)\sim 2.3\mu\Omega{\rm cm}$.  This is about 8 times smaller than the
measured value shown in Figs. \ref{fig:14} and \ref{fig:15}.  The small theoretical resistivity is a result of
the large theoretical value of $\omega_p$.   Alternatively, if we use the experimental
$\rho_1$ of Figs. \ref{fig:14} and \ref{fig:15}, and the BoltzTrap value of $\omega_p$, then $\lambda_{\rm tr}\ge 6$,
an unrealistic range.  All
known values of $\lambda_{\rm tr}$ are close to $\lambda$ \cite{Poole2000,Allen1987} 
which has never been observed greater than 2.
Large $\lambda$ may reflect small phonon frequencies, driven toward lattice instability, as is indeed found 
in computations \cite{Chan2013} for sc BP
when the pressure decreases toward 10GPa.  Large $\lambda$ would cause the superconducting $T_c$ to be much larger
than seen in BP.

Agreement in shape between measured $\rho(T)$ and Bloch-Gr\"uneisen theory is usually a good
confirmation of the applicability of the theory.  This would suggest that the measured value of $\rho(T)$ is somehow
too large by a factor $\sim$ 8.  But the similarity in magnitude and shape of the Li {\it et al.}
data to other experiments argue against this.  Also the downward curvature of $\rho(T)$ (seen by Okajima {\it et al.} 
and shown in Fig. \ref{fig:14} at higher
$T$) would be very unusual if the actual $\rho(T)$ were 8 times smaller than reported.
It should be mentioned that Wu {\it et al.} \cite{Wu2018} compute $\rho(T)$ from first principles.
The numerical values shown in Fig. 12 of their paper are only a factor $\sim 2$ smaller than experiment.
However, they apparently use a free electron choice of $(n/m)$, and values of $\lambda_{\rm tr}$
similar to those used here.  Therefore the plotted magnitude of $\rho(T)$ seem to have been incorrectly
enhanced by a factor $\sim 5$.

\section{speculations}

Electronic structure calculations \cite{Chan2013,Li2018} show that in the sc phase, harmonic phonons are not stable
unless $P>$20GPa.  This can have two interpretations.  
Either there is a DFT problem, and a correct harmonic theory would have stable phonons,
or else, more likely, DFT is correct about the instability of harmonic phonons, and  the sc phase is stabilized 
by anharmonic interactions when 10GPa$<P<$20GPa.
The sc phase is rare in nature.  P atoms have small masses and loose packing in the sc phase..
Thus one expects large zero-point vibrations.   
These factors likely require anharmonic interactions to be included in a correct zeroth order theory.
In such a situation, the large displacements can not only stabilize a harmonically unstable phase, but also alter
the electron-phonon coupling.  Coupling beyond first-order $(\partial U/\partial u_{\ell\alpha}) u_{\ell\alpha}$
will affect $\rho(T)$ in a way that alters the Bloch-Boltzmann theory from Bloch-Gr\"uneisen form.
Then probably one needs to invent a ``strongly coupled'' theory of lattice vibrations and their
interaction with electrons.  Numerous thoughts in this direction are available, for example, refs. 
\onlinecite{Kim2018,Ravichandran2018}.

\section{acknowledgements}

We thank Jianshi Zhou for stimulation, and Jinguang Cheng for help.  This work was supported by
 DOD-ARMY grant W911NF-16-1-0559. 

%



\bibliography{BPrho}

\end{document}